\documentclass[preprint]{aastex}
\usepackage{natbib}
\usepackage{graphicx}
 \usepackage{color}
 \newcommand{\ra}{$\rightarrow$}
 \newcommand{\rs}{$\mathrm{R_{\odot}}$}
 \newcommand{\kms}{$\mathrm{km\,{s}^{-1}}$}

\received{--}
\revised{--}
\accepted{--}
\slugcomment{To be submitted to ApJ}
\shorttitle{Fluxrope Formation Prior to CMEs}
\shortauthors{Patsourakos, Vourlidas \& Stenborg}
\begin{document}
\title{Direct Evidence for a Fast CME Driven by the Prior Formation and Subsequent Destabilization of a Magnetic Flux Rope}
\author{S. Patsourakos\altaffilmark{1}, A. Vourlidas\altaffilmark{2},
  G. Stenborg\altaffilmark{3}}  

\altaffiltext{1}{University of Ioannina, Department of Physics,
  Section of Astrogeophysics, Ioannina, Greece} 

\altaffiltext{2}{Space Sciences Division, Naval Research Laboratory, Washington, DC,
  USA}
\altaffiltext{3}{George Mason University, Fairfax, VA, USA}

%
%
\begin{abstract}
  Magnetic flux ropes play a central role in the physics of Coronal
  Mass Ejections (CMEs). Although a flux rope topology is inferred for
  the majority of coronagraphic observations of CMEs, a heated debate
  rages on whether the flux ropes pre-exist or whether they are formed
  on-the-fly during the eruption.  Here, we present a detailed
  analysis of Extreme Ultraviolet observations of the formation of a
  flux rope during a confined flare followed about seven hours later
  by the ejection of the flux rope and an eruptive flare. The two
  flares occured during 18 and 19 July 2012.  The second event
  unleashed a fast ($>$ 1000 \kms) CME.  We present the first direct
  evidence of a fast CME driven by the prior formation and
  destabilization of a coronal magnetic flux rope formed during the
  confined flare on 18 July.
\end{abstract}


\section{Introduction}
All currently available theories of CME formation predict that CMEs
are basically ejections of magnetic flux ropes.  This prediction is
largely confirmed by coronagraphic observations of CMEs in the outer
corona showing that the majority exhibits a clear flux rope  geometry
\citep[at least $\sim$ 40\%][]{av121}.  However, an intense debate exists on whether
such a magnetic topology exists before the CME onset or whether it is
formed during the CME eruption \cite[see for example the reviews
of][]{forbes00,klim01,chen11}. A conclusive answer to this question would represent
an important advance in our understanding of CMEs.  There are models of
CME initiation which form the flux rope once the CME is underway,
i.e. {\it on-the-fly\/}, while others require a {\it pre-existing flux
  rope} before the CME onset. A further division in the
latter models concerns the origin of the flux rope. It may be formed
either in the corona, {\it coronal flux rope}
\citep[e.g,][]{moore92,anti1999,amari2000,lynch08,vra08}, or in the
photosphere or low chromosphere, {\it photospheric flux rope}
\citep[e.g.,][]{mag01,manch04,gf06,archo08} after the emergence
of a twisted flux tube from the convection zone.

Significant progress into this problem has been achieved for CMEs
originating from quiet Sun (QS) regions, such as those associated with
Polar Crown Filaments. Observations in various spectral domains like
the White Light (WL) the extreme ultraviolet (EUV) and soft Xrays
(SXRs), has shown large-scale quiescent cavities going through a
quasi-static rise phase, with a duration of up to several days, which
could eventually erupt giving rise to CMEs
\cite[e.g.,][]{engv89,huds99,koutch04,gib06,su10,regn11}. Cavity densities are typically
lower and their temperatures are higher than the background
corona values \cite[e.g.,][]{full08,vasq10,kuc12,reeves12}. It is widely
acknowledged today that the observed cavities (or at least part of their
 cross section) correspond to a magnetic flux rope seen
edge-on. Further evidence for a flux rope topology in QS cavities
comes from polarization signals of the coronal magnetic field
\citep{dove11}, from swirling motions observed in these structures
\citep{wang10} and from concave-upward structures in CME
cavities \citep{plunkett00, rob09, av121}.  Therefore, the pre-existing
flux rope scenario seems very viable for CMEs originating in the QS.
Finally, prominences either quiescent
or eruptive often exhibit helical structures,
which represents a strong indication of a flux rope
topology \cite[e.g.,][]{vra91,romano03,rust05,willia05,chen06}. Note that
is widely believed today that helical prominences correspond
to only the lower parts of flux ropes, where dense material
is collecting.

However, the situation is  unclear for impulsive CMEs originating
in active regions (ARs). This is due to several 
physical and geometrical factors. First, the prevalence of higher
magnetic field orders in ARs compared to QS implies that the
CME-related structures (cavity, flux rope, etc) will be both smaller
in size and lower-lying in ARs compared to the QS. This would make it
difficult to identify any pre-existing flux rope, especially when we
consider the line-of-sight (LOS) interfence from the EUV emission of
the myriads of loops lying at more or less random orientations in the
low corona.  Second, impulsive CMEs in ARs evolve at relatively short
time-scales because of the smaller spatial scales and higher
Alfv\'{e}n speeds in ARs compared to the QS \citep[e.g.][]{vra07}. The rapid evolution would
make it extremely difficult to discriminate between pre-existing and
on-the-fly formed flux ropes. Indeed, recent high cadence observations
of impulsive CME onsets have placed strong constraints on the initial
sizes and timescales of potential flux ropes
\citep[e.g.][]{pvk,pvs,av12}. These limits were inferred from observations
of the formation and evolution of CME cavities in the EUV, and showed
that CME flux ropes could be initially very small (radius $<$ 0.01
$\mathrm{R_{\odot}}$) and low-lying (height $<$ 0.1
$\mathrm{R_{\odot}}$) and could evolve in short time scales of the
order of 1-2 minutes. Third, impulsive CMEs from
ARs are always associated with significant plasma heating in the form
of a flare. According to the standard CME-flare {\it CSHKP}
model \citep[e.g.,][]{carmi64,stur66,hira74,kop76}, one part of the
reconnected magnetic flux under the erupting CMEs is directed towards
the flare loops and the other part is added to the erupting flux rope.
Hence, the erupting flux ropes will be (at least initially) substantially
heated. Therefore, CME flux ropes should appear in hot,
flare-like, EUV wavelengths which were not routinely observed at high
cadence until recently. Taken together, these four reasons explain
why flux ropes, whether pre-existing or not, are so elusive in impulsive CMEs.

Non-linear force free magnetic field extrapolations and flux rope
insertion methods reveal magnetic field distributions pertinent to
flux ropes in ARs which could eventually erupt
\citep[e.g.,][]{canou09,sav09}.  Although quite valuable, such
calculations rely upon the observed photospheric magnetic field
which does not change significantly over large areas during
eruptions. Therefore, such methods may not be particularly
helpful for tracking the (presumably) rapidly changing coronal magnetic
fields during CMEs. This limitation together with the
frozen-in property of coronal plasmas to the coronal magnetic field
make rapid, narrow-band multi-thermal EUV imaging going all the way
from flare down to transition region temperatures a very powerful tool
to study rapid changes in the coronal field during CME onsets.

Strong, but indirect, evidence for pre-existing flux ropes comes from
EUV and SXR observations of the formation and eventual eruption of
sigmoids \cite[e.g.,][]{aura99,canf99,vra03,green09,tri09,liu10,green11,huang11}.  EUV
and SXR sigmoids are interpreted as disk signatures of flux ropes
viewed from above. Since sigmoids are optimally observed from above,
projection effects may enter into the interpretation of the
observations (e.g.  structures at different heights may appear
connected in projection). SXR observations of an eruptive
flare by SXT showed the formation of an oval-shaped   structure, highly suggestive
of a flux rope core,
at the onset of the impulsive CME acceleration and simultaneously with the appearance of the X-ray
flare loops \citep{vra04}. Moreover, SXT and XRT observations
showed evidence of cusp-shaped loops
forming under  the erupting flux which sometimes
has a concave upwards V-shape \cite[e.g.,][]{tsu92,tsu97,sava10}. All these features
are predicted by the standard solar eruption model. 
More recently, EUV observations in flare
temperatures ($\approx$ 10 MK) of CME onsets at or close to the solar
limb showed the formation of a magnetic flux rope a few minutes before
the onsets of the associated CMEs and flares \citep{cheng11,zhang12},
with its upper part resembling a plasmoid structure \citep{reeves11}.

Finally, the question of whether the flux rope forms before or during
the eruption has broad implications for the CME initiation
mechanism(s). The answer will determine whether the eruption process
is ideal (prior flux rope) or resistive (on-the-fly flux rope).  As we
argued, the question was very difficult, if not impossible, to answer
until recently because of low cadence and sensitivity, and lack of
observations in appropriate hot EUV lines.  The availability of high
cadence, multi-wavelenth EUV observations from the Atmospheric Imaging
Assembly \citep[AIA;][]{aia} aboard the \textit{Solar Dynamics
  Observatory\/} (SDO) since 2010 and the multiviewpoint observations
from the EUV Imager \citep[EUVI;][]{euvi} in the Sun Earth Connection
Coronal and Heliospheric Investigation \citep[SECCHI;][]{secchi}
imaging suite aboard the \textit{STEREO\/} mission have greatly
improved the situation.

In this paper,we directly address the question of the pre-existing
fluxrope for a fast ($>$ 1000 \kms) CME that erupted on 2012 July
19. We take advantage of detailed EUV observations, spanning several
hours, of its source region before the eruption.  The
multi-temperature coverage from the AIA images allows us to detect and
follow (thermally and kinematically) the formation and evolution of a
very clear flux rope  structure on July 18 at $\approx$ 22:20 UT
during a confined flare. The EUVI observations allow us to reconstruct
its three-dimensional (3D) morphology. The flux rope finally erupts on July
19 at around 05:20 UT creating the fast CME. We present the EUV
observations and analysis of the flux rope in Section 2 and the inferred 3D
structure in Section 3. In Section 4, we discuss the evolution of the
flux rope and the implications for CME intiation theories. We conclude in
Section 5.

\section{Observations and Data Analysis}
We analyzed EUV images of the low corona (1-1.3 \rs and 1-1.6 \rs,
respectively) from the \textit{SDO\/}/AIA and \textit{STEREO\/}/EUVI,
and WL images of the outer corona (2.2-6 \rs) from the Large Angle and
Spectrometric Coronograph \citep[LASCO;][]{lasco} C2 coronagraph on
board the \textit{SOHO\/} mission.

We used AIA images (level 1.5) recorded in narrow-band channels
centered at $\approx$ 304, 171, 193, 211, 335, 94 and 131 \AA, which
have peak responses at temperatures of $\approx$ 0.05, 0.6, 1.6, 2.0,
2.5, 6.3 and 10 MK respectively; EUVI images from the 195 \AA \,
channel having a peak response for $\approx$ 1.6 MK were also used.
In the rest of the paper we will refer to any given channel by simply
supplying the wavelength of peak response: for example 304 \AA \,
channel will be referred to as 304.  The signal in 94 and 131 is
dominated by multi-million plasmas only when intense heating, usually
associated with flares, takes place. As we will see, the AIA
capability to obtain narrow-band images of ultra-hot plasmas is a
decisive factor in understanding the eruption process in this
event. Under quiescent conditions the signal in 94 and 131 is
dominated by cool emissions formed below 1 MK. Moreover, total
brightness images of the corona taken by LASCO on SOHO were
analyzed. To reduce data volume, we used AIA images with a reduced
cadence of 1 minute. Inspection of full cadence movies during the
$\sim 9$ hour window of our investigation showed that the one minute
cadence was sufficient to resolve the various dynamics. The cadence of
the STA 195 and LASCO images was 5 and 12 minutes, respectively.  Pixel
sizes for AIA, EUVI and LASCO C2 images are 0.6, 1.6 and 12 arcsec
respectively.  To enhance the image contrast in the corresponding EUV
images we have processed the original data with wavelets
\citep[][]{stenb08} to bring out their fine structure; we essentially
subtracted from each frame a ``background'' frame resulting from a
wavelet-filtered version of the frame amplifying only the low spatial
frequencies (i.e., enhancing tha large-scale structure).  However,
every flux measurement was carried out on the original data.

Our target was NOAA active region (AR) 11520.  This AR was
particularly active and hosted several flares and CMEs. We hereby
focus on the time interval from $\approx$ 21 UT on 2012 July 18 until
06 UT on July 19 when AR11520 hosted a {\it confined} C4.5 flare (peaked at
$\sim 22$ UT on July 18) and an {\it eruptive} M7.7 flare (peaked at $\sim$ 06 UT
on July 19).  The eruptive flare was associated with a
WL CME observed later on by LASCO C2. Figure 1 summarizes the events
as seen from different perspectives.  During the period of interest,
11520 was located at the West limb as seen from AIA (Earth) and at almost $30^\circ$ East off the central meridian as viewed from EUVI on STEREO
Spacecraft A (EUVI-A, hereafter), which was $120.6^\circ$ ahead of the Earth. Thus
we had both edge-on (AIA) and face-on (EUVI-A) observations of the
target AR.  Figure 1 also
shows the photospheric magnetic field distibution around AR11520
on  July 12, when the AR crossed the central
meridian passage as viewed from Earth using the Helioseismic and
Magnetic Imager (HMI; \cite{hmi}) on  \textit{SDO\/}.  The complexity of the
photospheric magnetic field is obvious. A curved neutral
line (NL) shape can be inferred from the magnetic field
distribution. Proxies of the NL shape and extensions during the period
of interest could be inferred by the inspection of dark filament
material images of the target AR as seen from above in 304 and 195
EUVI-A images. The NL was quite long and complex and consisted of several
approximately linear segments forming a mirrored ``?''  shape
starting from the eastern end of S3, continuing along
S1 and S2 and
extending farther southward.
The AIA LOS was aligned
with horizontal element S1, vertical element S2 was running north-to-south
parallel to the West limb as seen from the Earth, and element S3 was
well behind the West limb and therefore was invisible
from Earth.  Note the sheared shape of the NL at the junction
connecting S1 and S2.  This was probably related to a rotating spot
dragging and shearing the magnetic field at this location. This
feature may have significant implications for the AR evolution.  The
footpoint alignment between the AIA LOS and element S1 was an important
factor for observing the flux rope structure, as we will see later.

In the following subsections we organize our observations into three
distinct phases. Their synthesis into a cohesive physical and geometrical scenario
comes in Sections 3 and 4.

\subsection{Formation of the Flux Rope and Failed Eruption}
The first phase consists of a confined flare on July 18 accompanied by
the formation of a very clear flux rope structure.  These events are
readily observed in the 131 online movie
(movie1.mp4). Figure~\ref{fig:conf} contains several snapshots from
this movie.  Starting at around 22:00 UT, we observe the rise of a
narrow structure above the western limb (Figure
\ref{fig:conf}b). Eventually, the structure exhibits a core, with an
elliptical cross-section, and several 'legs' threading the core and
connecting only on the southern side (Figure~\ref{fig:conf}c).  In
addition, cusp-like loops appeared underneath the core of the
structure (Figure~\ref{fig:conf}d). The structure bears a stricking
resemblance to cartoon depictions and MHD model snapshots of
magnetic flux ropes.

Just 25 minutes after its start, the structure stops
rising. Concurrently with these motions, a C4.5 flare from the same
active region is taking place. The bright cusp-like loops are
associated with the flare, as a quick inspection of the GOES SXI images
reveals. No permanent dimmings are detected across the various AIA EUV
channels over the area covered by the rising structure. No CME is seen
in the LASCO C2 image, expect for an evanescent compression
wave. These observations are consistent with the 'failed CME' events
identified in \cite{av10}. The wave is very likely a compression wave
launched by the ascent of the structure (piston-driven). Although it remains a possibility, it is unlikely that the wave could be launched by a flare-related pressure pulse because the flare heating occurs after the rise phase (Figure~\ref{fig:confht}).
As we will see in
the next section, we find direct evidence of cooling of the flare
plasmas entrained within the structure. Therefore, these observations
suggest strongly that we are dealing with a failed eruption and consequently with a {\it
confined flare.}

The most interesting and novel aspect of these observations is that
the ``core+legs'' structure formed during the confined flare
represents a clear example (possibly 
the clearest example in the literature so
far) of a magnetic flux rope. We believe that this is the case for the following reasons:
\begin{enumerate}
\item It exhibits a {\it coherently evolving} large-scale structure
  enclosing {\it intertwined} threads (see for example the high-pass
  filtered image in Figure~\ref{fig:conf}d). The coherency implies
  that the observed structure is a single macroscopic structure (an
  essential element for a magnetic flux rope interpretation) and not
  the fortuitous alignment, along the LOS, of the expansions of
  individual loops. Compare, for example,
  with the ``independent'' loop expansions observed prior to CME
  cavity formation \citep[e.g.,][]{pvs,pvk}. The internal fine coiling structure is a major
  ingredient of any magnetic flux rope.
\item It is formed by {\it magnetic reconnection}.  The observations
  show that the 131 ``core+legs'' structure is appear concurrently
  with the cusp-shaped loops underneath, which also stretch and rise
  during the ascent of the structure. This is probably the most
  straightforward evidence that magnetic reconnection is forming a
  magnetic flux rope. According to the standard model of solar
  eruptions, the erupting flux generates coiled field lines which
  become part of a newly-formed (or add new flux to a pre-existing)
  flux rope. Interestingly, our observations show that the core is
  formed progressively through the continuous addition of new outer
  layers. This is the expectation from reconnection adding new flux to
  a flux rope \citep[e.g.,][]{lin04}.  The cusp underneath the
  erupting flux represents the boundary between the most recently
  reconnected magnetic field lines which were either retracted
  downwards to form the post-eruption loops (i.e. the flare loops) or
  upwards to become part of the flux rope. A current sheet is expected
  to form between the tip of the cusp and the bottom of the flux rope.
  Finally, we note that the cusp and the narrow lower part of the flux
  rope core resemble an "X". This is strongly suggestive of the
  formation of a {\it coronal X-point}, in 2D, or more generally of a
  quasi-separatrix layer (QSL), in 3D
  \citep[e.g.,][]{aula10,sav12}. QSL layers separate domains of
  distinctively different magnetic topology, like flux rope and non
  flux rope fields in our case. Strong currents are expected to
  develop there.  More importantly, QLSs represent regions where
  magnetic reconnection can occur easily.
\item It is very {\it hot} ($\sim 10$ MK) because it is initially
  visible only in the 131 channel. The temperature map in
  Figure~\ref{fig:conf}f corraborates that the structure attains
  temperatures of around 10 MK during the confined flare. The map is
  calculated from a Differential Emission Measure (DEM) analysis of
  the AIA fluxes in the six EUV coronal channels (94, 131, 193, 171,
  211, and 335) taken almost simultaneously during the confined
  flare. We use the methods and software described in
  \citet{demma}. Essentially, a Gaussian DEM is simultaneously fitted
  to the observed fluxes in each pixel. Our fitting searches for 
  DEM peak temperatures in the
  range 0.5-25 MK.  The displayed temperature map in
  Figure~\ref{fig:conf}f corresponds to temperature of the peak best-fit DEM.  
  The flux rope reached a maximum temperature of
  $\approx$ 12 MK. Since the "core+legs" 131 structure gets, at least
  initially, as hot as the flare loops underneath, their formation
  is likely sharing a common physical origin. This must be the reconnection
  in the current sheet below the rising flux rope. It
  heats the plasma in field lines which either become part of the
  flaring loops or of the rising flux rope.
\item The flare brightenings {\it map along} the NL. This is
  easily determined by checking the EUVI-A images above the target AR
  (Figure~\ref{fig:sta}).  Comparing the shape and orientation of the
  NL before the eruption (Figure~\ref{fig:sta}a) with the brightenings
  taking place during the confined flare (panels b-c), we see that the
  brightenings run almost parallel to the horizontal segment
  ('S1'). These ribbon-like brighenings correspond to the footpoints of
  the field lines energized by the flare.  The pattern is widely
  consistent with flux rope formation from a sheared arcade: arcade
  fields with footpoints almost perpendicular to the NL are
  transformed (through slow shearing motions, for example) to coiled
  flux rope field lines connecting distant points along the NL. The final
  footpoints of these flux rope field lines are arranged at opposite
  ends of the NL running more or less parallel to it.
\item It exhibits a {\it ``half-loop'' topology}.  A striking feature
  of the ``core+legs'' 131 structure is that only the southward part
  of the legs is visible (e.g., Figure~\ref{fig:conf}c-d). There is no
  obvious extension of these legs to the north. On the other hand, the
  legs of the post-eruption loops, underneath the structure, are fully visible on either side of
  the "X" point. This discrepancy is explained in Section~3 and
  provides us with a very strong indication that we are dealing with a
  twisted three-dimensional flux rope structure.
\end{enumerate}

One may be tempted to describe the hot core observed in 131
as a ``blob'' or a ``plasmoid''. Note that both terms arise
from 2D or 2.5D depictions and imply structures partially or fully
detached from the solar surface which could eventually escape the
instrument's field of view. However, the hot core
is attached to the surface via the legs discussed above
and does not escape from the Sun.

We emphasize here that the formation of the flux rope would have gone
largely unnoticed without the hot channel observations. The warmer
channels like 171, 193, or 211 (i.e., the 211 on-line movie, movie2.mp4) show only 
expansion and rise of loops overlying the 131 flux rope structure in phase
with the expansion of the 131 flux rope structure. 

To deduce the flux rope kinematics and compare it to the various emissions we
manually trace the height of the flux rope front in the 131 images
(Figure~\ref{fig:confht}). The upper panel displays the height-time
($h-t$) measurements (square boxes).  We assign a conservative error
of five AIA pixels (0.003 \rs) to every measurement.  The $h-t$ data
are smoothed first to reduce small-scale fluctuations.  We use a
smoothing cubic spline scheme \citep[e.g.,][]{weisb05} which minimizes
a function consisting of the sum of a ${\chi}^{2}$ fit of the data
with a cubic spline plus a penalty function proportional to the second
derivative of the cubic spline. Five knots are found to minimize the
Akaike Information Criterion (AIC) and the Bayesian Information
Criterion (BIC). Both AIC and BIC are standard measures of the
relative goodness of a statistical model and supply a means of model
selection \citep[e.g.,][]{liddle07}.  
The flux rope starts to slowly rise at around
22:00~UT until around  22:20~UT, when
it reaches a constant height (i.e., it "stops"
at the corresponding height).
Next, we
derive the evolution of flux rope speed and acceleration by taking the first
and second numerical time derivatives of the smoothed $h-t$
measurements (solid and dashed black lines, respectively in
Figure~\ref{fig:confht}, middle panel). We then perform 1000
Monte-Carlo simulations of the (assumed) Gaussian $h-t$ uncertainties
to derive the 1-$\sigma$ point-wise uncertainties for the velocity and
acceleration (red and blue curves in Figure \ref{fig:confht}, middle
panel).

During the impulsive rise phase the velocity reaches a peak value of
$\approx$ 60 $\mathrm{{km}{s}^{-1}}$ and then decreases to $\approx$ 0
$\mathrm{{km}{s}^{-1}}$.  The flux rope undergoes an asymmetric short-lived acceleration pulse with a duration of $\sim 12$ few minutes. Note
we are "missing" part of its rise phase, i.e. the acceleration does
not start from zero. We attribute this to the abrupt appearance and
rise of the observed structure in the first frames used in our
measurements. We verified that this behavior was
not due to the 1-minute cadence images we used in our analysis:
we were not able to see any significant
change  that could be measured with some confidence before
21:59:33 UT (i.e. our first measurement point) 
even when browsing the full 12-s 131 data.  
The bottom panel of Figure \ref{fig:confht} contains
normalized curves (to their respective peak values) of: (i) GOES 1-8
\AA\ SXR light curve (solid line), (ii) its temporal derivative (as calculated
from averages over 10 full resolution temporal pixels;
dashed line) which is a proxy of the Hard Xrays (HXRs) and thus a
metric for the energy release rate due to flare reconnection, and
(iii) the 131 light curve (dash-dot line) integrated within a box
containing the flux rope as shown in Figure \ref{fig:conf}d. We focus on
light curves from a small area containing the flaring region to derive
a more accurate estimate for the true onset time and rise rate of the
associated flare. Generally speaking, the averaging over the entire
solar disk of the GOES measurements may lead to a delayed flare onset
times and/or shallower rises.  Comparing now the middle and lower
panels of Figure \ref{fig:confht} we find that the HXR proxy exhibits
a couple of short-lived pulses with the first reaching its peak
slightly {\it after} ($<$ 1 minute) the peak of the acceleration.  The
observed flare - flux rope dynamics largely conform with the well-known
synchronization between flare emissions and CME acceleration
\citep[e.g.,][]{zhang01,zhang04,mari07,temme10,beinsxr}.  For example,
the recent extensive statistical study of \citet{beinsxr} found for a
set of 95 events that: (i) CME acceleration starts {\it before} the
SXR flare onset (75 \% events) and (ii) the time delay between the
peaks of the CME acceleration and of the SXRs temporal derivative
occur within $\pm$ 10 minutes (81 \% of events).  Our event clearly
falls within these limits and delays.
  
Further insight into the formation and the subsequent evolution of the
flux rope can be obtained by comparing its size evolution with the
dynamics of the associated flare. We manually
select nine points outlining the flux rope core (avoiding the legs)
for several times in 131 and then fit an ellipse to
the selected points.  Figure~\ref{fig:radius} shows the temporal
evolution of the derived flux rope minor radius and area.  Given the almost radial
path followed by the flux rope, the minor (major) axis of the fitted
ellipses corresponds to the lateral (radial) extent of the flux rope
core.  The radii and areas are first corrected for the
instrumental resolution width (Table 7 in \cite{aia1}).  

Several remarks from Figure~\ref{fig:radius} are now in order. First,
the initial size of the flux rope core is very small (e.g. initial minor axis
radius of 4.7 Mm or 0.006 \rs).  Second, the minor axis and area of
the flux rope core, and hence its major axis, undergo a short-lived period of
strong growth starting at around 22:06 UT and lasting for 13 and 17
minutes, respectively.  The two-phase development of the flux rope minor axis
largely coincides with a two-phase activity (22:08 - 22:11~UT and
22:15 - 22:16~UT) seen in the GOES SXR derivative. The energy release
rate, peaks two minutes {\it before} (22:16~UT) the flux rope core reaches its maximum
size ($\sim 22:18$~UT). These timings strongly suggest that flare
reconnection was responsible for the flux rope formation. This is
corroborated further by the fact that the flux rope growth ceases
within 2-3 minutes of the SXR flare peak.  After $\approx$ 22:23 UT,
the flux rope maintains an almost constant size, which implies that the flux rope
formation is completed by this time. The flux rope also reaches its peak
height at this time (upper panel of Figure \ref{fig:confht}).

The initial height ($\approx$ 27.6 Mm; Figure \ref{fig:confht})
and minor axis  length $\approx$ 4.7 Mm Figure \ref{fig:radius})
of the flux rope clearly indicate that the structure was formed
at coronal heights. Moreover, the flux rope structure readily attained flare temperatures.
This argues in favor of the formation of a {\it coronal flux rope}.

\subsection{Quasi-Static Evolution and Cooling of the Flux Rope}

Once the confined flare ends at around 23~UT on July 18, the flux rope
structure undergoes cooling. Starting at the same time and lasting for
several hours (until around 04:00~UT on July 19) the flux rope and overlying
loops begin a phase of slow rise and expansion.

The cooling of the flux rope structure evolves coherently across the
different AIA channels after the end of the confined
flare. Essentially the hot 131 flux rope starts to appear sequentially in AIA
channels with decreasing characteristic temperatures.  An example of
this evolution is shown in Figure~\ref{fig:cool} where we see the flux rope
core and legs appearing in the different AIA channels at different
times. The core emission is stronger in the hotter channels (94 and
335) while the legs are better seen in the warmer channels (211, 193,
171, 304). Indeed, the flux rope core was visible for several hours in the
335 channel, even after the end of this phase at $\sim$ 04:00~UT on
July 19, which suggests that its temperature did not drop below the
characteristic temperature of 335 ($\sim 2.5$ MK).  The elevated
temperature likely explains the lack of a sigmoid in the EUVI-A 195
images (characteristic temperature of 1.3 MK) above the source AR.
Moreover, the cooling of the flux rope progressed from its interior to
its exterior, i.e. in the same sense to the heating of the flux rope  during its
formation.  These patterns are compatible with magnetic reconnection
during the confined flare adding new magnetic flux to the
flux rope \citep[e.g.,][]{lin04}. For example,the parts of
the flux rope formed earlier (the inner parts) would then cool earlier to a given temperature.
The cooling of the flux rope structure leads to a very important conclusion: the magnetic
structure of the flux rope was maintained for several hours after its formation during the
confined flare.

The cooling of the flux rope can be also
appreciated by inspecting the light curves 
of different AIA channels intergrated within a box containing
the flux rope and shown in Figure~\ref{fig:conf}d.
The AIA light curves, together with the GOES 1-8 \AA\
light curve, are displayed in Figure \ref{fig:lcurves}.
The plot shows emission peaks in the various channels,
early in the plotted timeline,
with an ordering as a function of the temperature
of peak response in each channel: 131\ra94\ra335\ra211\ra193\ra171\ra304.
Additionally, we notice another emission peak in 304 occurring
before the higher temperature peaks. This is probably emission from the footpoints of the
flaring loops, which preceeds the emission from their coronal
sections. 

While the flux rope cools, the AIA movies show evidence of a slow
rise and expansion of the flux rope and its overlying loops. This is
particularly evident in the warmer channels.  The expansion continues
for several hours, until about 04:00 on July 19.  To better visualize
and quantify the slow expansion and rise phase, we create a stack plot
of the temporal evolution of the intensity, for several AIA channels,
along the path shown in Figure~\ref{fig:conf}d. The path contains the
flux rope and overlying loops. A sample of the resulting stack plots
is shown in Figure \ref{fig:jmaps}.  First, we note a series of
low-lying bright streaks around the time of the confined flare. These
correspond to the flux rope observed in different channels.  Second,
we note a series of almost linear intensity tracks, with positive
(hence rising) slopes, starting at different heights above the flux
rope (dashed lines).  These tracks can be seen in different channels and last from
$\approx$ 00:00 to 04:00. They correspond to the slow rise of the
overlying loops. The slopes of these linear tracks yield speeds in the
range 0.5-2.0 \kms. Because the magnitude of these speeds is a small
fraction of the characteristic speed in the AR coronal core ($\sim 1000$ \kms), the
observed rise and expansion can be described as a quasi-static
process. 
In addition, the solar rotation is too slow to explain the
observed rise during the few hours we consider here.
During this interval, and in tandem with the cooling
of the flux rope and slow rise of the overlying loops,
we also observe evidence of activities taking place
at and around the flux rope core, including apparent
displacements of its legs. The latter implies some sort
of magnetic field reconfiguration.

This quasi-static rise of the overlying loops can explain a steady,
slow decrease in the intensity of the warmer channels (e.g., 211, 193,
171) which starts at around 02:00~UT (see Figure
\ref{fig:lcurves}). It is simply due to the slow evacuation of loops
from our selection box.

\subsection{Destabilization and Eruption of the Flux Rope}

Starting at around 03:00~UT, the system enters into a new phase
leading to the destabilization and eruption of the flux rope, a strong
flare and eventually a fast CME. We post an on-line movie (movie3.mp4)
showing a composite of 335-131 images from this phase. Figure
\ref{fig:erupt1} contains several snapshots from this movie.

\begin{figure} 
\includegraphics[scale=0.8]{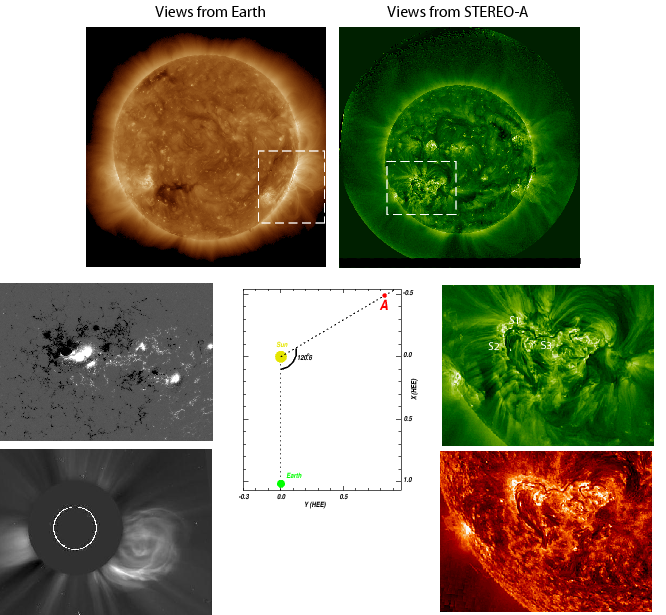} 
  \caption{Context information for the flare and CME events in this study.
Middle panel: Relative positions of Earth, Sun and STA during
our observations. Left panels: Earth-based views
of the source AR. (top) AIA 193 image  on 2012 July 18. (Middle)
HMI magnetogram on 2012 July 12, close
to the central meridian passage of the AR as seen from the Earth. (Bottom) LASCO C2 image
showing the CME associated with the eruptive flare on July 19.
Right panels: Views from STA on 2012 July 18. (top) 195 full disk image. (Middle) detail of the source AR in 195 and 304 (bottom). The boxes in the upper panels contain the source AR.
}\label{fig:context}
\end{figure}

\begin{figure} 
\includegraphics[scale=0.9]{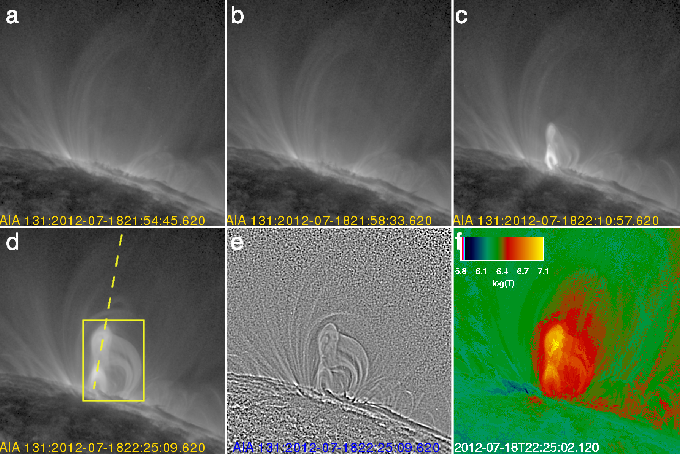}
\caption{Representative snapshots in the 131 channel of the confined
  flare. (a) The pre-event configuration, (b) the start of the flux
  rope rise, (c) the full development of the flux rope and (d) the
  `flux rope reaches a stable height and size.  The images are scaled
  logarithmically.  Panel (e) shows a wavelet-enhanced version of
  panel (d) to highlight fine structure within the flux rope. (f) Temperature
  map at the time of panel (d). The field of view is 600$\times$ 600
  $\mathrm{{arcsec}^{2}}$. The images have been rotated so that
  East-West represents the vertical direction. The box and the dashed
  line overplotted on panel (d) depict an area and path used in the
  construction of Figures \ref{fig:lcurves} and \ref{fig:jmaps},
  respectively.}\label{fig:conf}
\end{figure}

\begin{figure}
\includegraphics[scale=0.9]{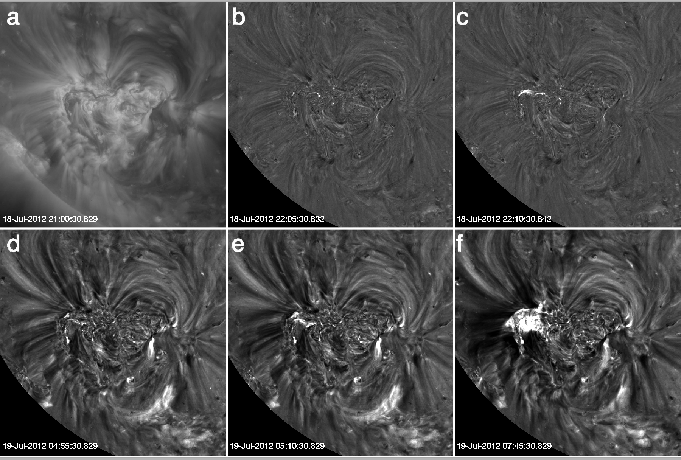}
\caption{Evolution of AR 11520 during 2012 July 18-19 as seen by
  EUVI-A 195.  (a) Direct image (log-scaled) of the scene. (b-c) and,
  (d-e) show the evolution of the flare brightenings during the
  confined and eruptive flares, respectively.  (f) the
  post-eruption arcade resulting from the eruptive flare. Panels (b-f)
  are base-ratio images (linear scaling from 0.2-3) of the EUVI-A 195
  images relative to the image in (a).  The images have been
  differentially rotated to the time of the base image before taking
  the ratios. Brighter gray level correspond to higher intensities or
  ratios.}
\label{fig:sta}
\end{figure}

\begin{figure}
\includegraphics[scale=0.8]{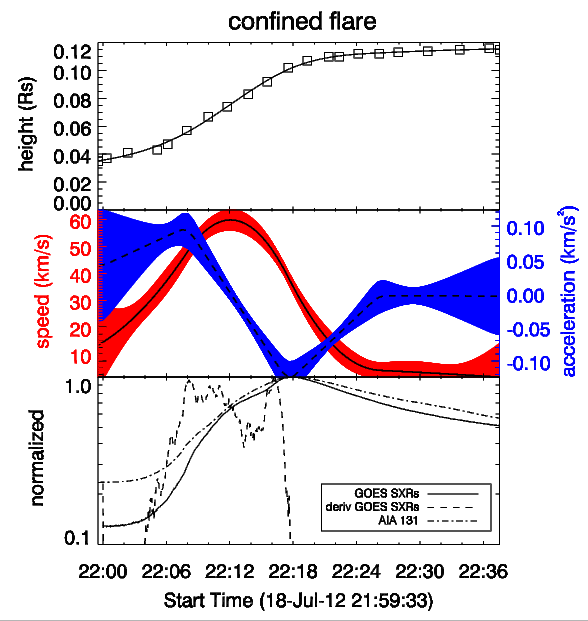}
\caption{Confined flare.  Top: temporal evolution of the flux
  rope height based on measurements in AIA 131 \AA \, (squares). The associated uncertainties are too small to be seen in
  this scale.  The heights resulting from the application of a
  smoothing cubic spline to the original measurements are plotted with a
  solid black line. Middle: first (velocity)
  and second (acceleration) temporal derivatives of
  the smoothed height measurements (solid and dashed black lines
  respectively). Point-wise 1-$\sigma$ uncertainties in the velocity
  (red) and acceleration (blue) from 1000 Monte-Carlo simulations of
  the (assumed) Gaussian $h-t$ uncertainties.  Bottom:
  normalized light
  curves of the GOES 1-8 \AA \, flux (solid line), its temporal
  derivative (dashed line) and the 131 light curve in the box
  encapsulating the flux rope structure in Figure
  \ref{fig:conf}d (dashed-dotted line).} \label{fig:confht}
\end{figure}

\begin{figure}
\includegraphics[scale=0.9]{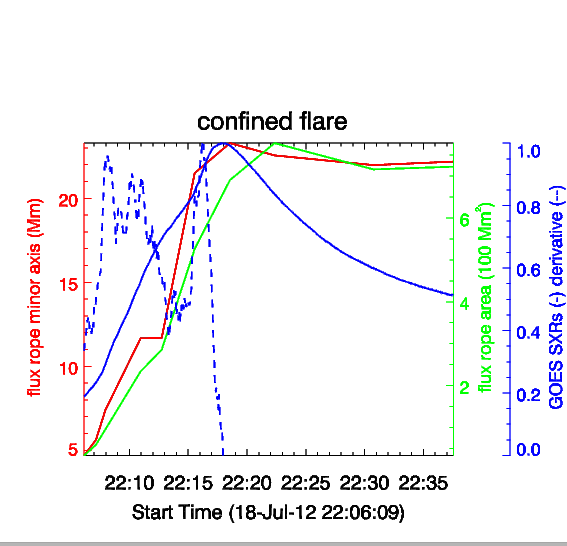}
  \caption{Evolution of the flux rope (full) minor axis (red) 
and of its area (green)
from ellipse fittings of the 131 flux rope core
for the confined flare. Normalized to their
peak values GOES SXRs (blue solid line)
and its temporal derivative (blue dashed line)
 are also displayed.} 
\label{fig:radius}
\end{figure}

\begin{figure} 
\includegraphics[scale=0.9]{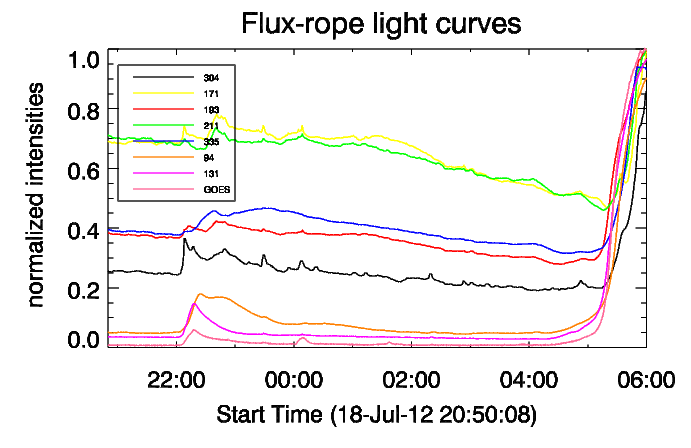} 
  \caption{Light curves in various AIA  channels within the box encapsulating
the flux rope structure as shown in Figure \ref{fig:conf}d and in
the GOES 1-8 \AA \, channel. All curves are normalized to their
respective peak values.}
\label{fig:lcurves}
\end{figure}

\begin{figure}
\includegraphics[scale=0.9]{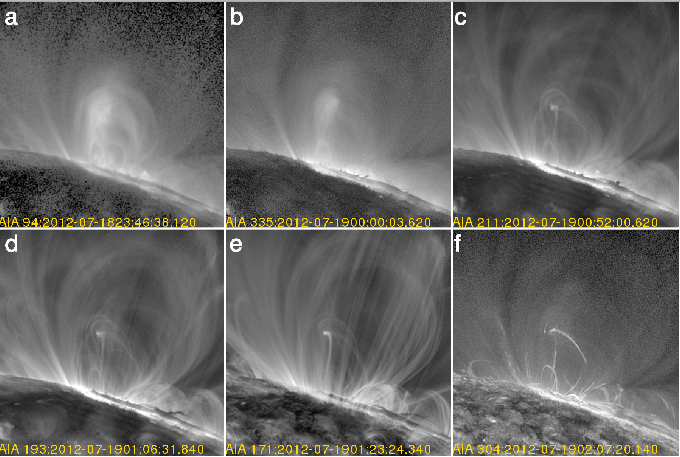}
  \caption{Cooling of the flux rope structure after the confined
flare as seen in various  AIA channels.  The field of view is  600$\times$ 600
$\mathrm{{arcsec}^{2}}$. Images have been rotated so that their
vertical dimension corresponds to the East-West direction.
  } \label{fig:cool}
\end{figure}

\begin{figure} 
\includegraphics[scale=0.9]{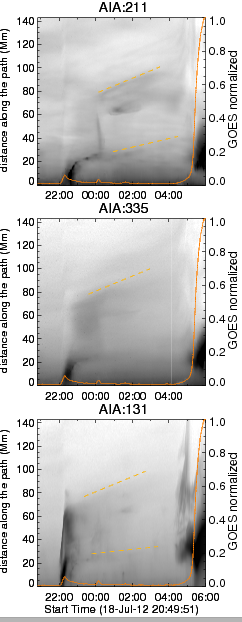}
\caption{Stack plots of the intensity along the path shown in Figure
  \ref{fig:conf}d.  Plots in 211, 335 and 131 are displayed in the
  top, middle, and bottom panel, respectively.  Time increases from
  left to right and distance along the path increases from the bottom
  up. The intensity in each point is the average over a 5-pixel wide
  slit running perpendicularly to the path.  The color table is
  reversed, i.e. darker shades correspond to higher intensities. The overplotted
dashed lines point to a sample of the slow rise tracks.}
\label{fig:jmaps}
\end{figure}

\begin{figure}
\includegraphics[scale=0.9]{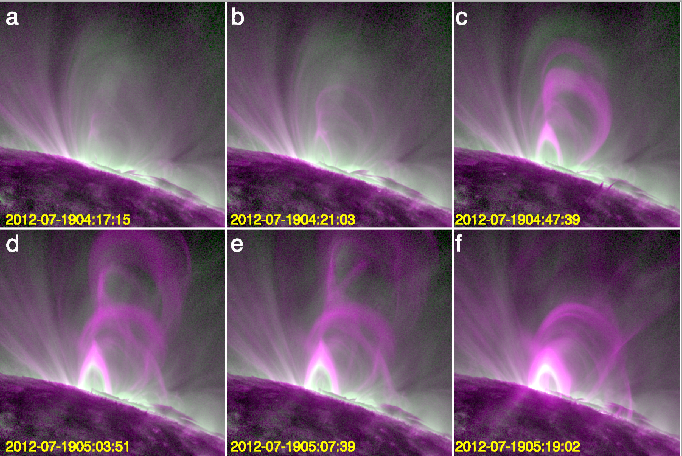}
  \caption{131 (red) and 335 (green) color composites
before and during the eruptive flare on 2012 July 19. 
The time tags correspond to the 335 images. The corresponding 131
images were taken 6 sec later. The field of view is 600$\times$ 600
$\mathrm{{arcsec}^{2}}$. The images have been rotated so that their
vertical dimension corresponds to the East-West direction.
  } \label{fig:erupt1}
\end{figure}

Starting at around 02:47~UT, the 131 images show a cusp brightening
below the initial flux rope along with the appearance of "half-loop"
structures, similarly to what was observed during the confined flare.
At the same time, the 335 flux rope core is rising slowly. These motions
become more pronounced from 03:57~UT onwards, when the cusp and
"half-loops" start to grow faster and the 335 flux rope rises at
a faster pace (Figure~\ref{fig:erupt1}).  As discussed earlier, the existence and
development of hot cusp structures points to magnetic reconnection
taking place above these structures. We believe that this process is
adding new flux around the erupting flux rope core.  This can be seen in
panels (e) and (f) of Figure \ref{fig:erupt1} where we observe the 335
flux rope core "sitting" on top of a concave upwards V-shaped 131
structure. The latter may be "nested" around the flux rope core via
magnetic reconnection above the cusp
\citep[e.g.,][]{lin04}.  The flux rope core and leg system continue
to rise and the flux rope core exits the AIA FOV at 05:07~UT.
Note here that the SXR levels were relatively low (less
than $\approx$ B3 of the GOES scale) during these
evolutions. The associated flare is still at its gradual rise phase
when the flux rope core exits the AIA FOV.  
Around 05:36~UT a WL CME emerges in the LASCO C2 coronagraph. The CME is
a typical flux rope CME (see for example the bottom left
image of Figure \ref{fig:context} and \cite{av121} for definitions). 
The CME front exits the C2 FOV at around 06:00~UT when the associated M7.7 GOES class flare
reaches its peak. We thus conclude that the flux rope rise described
above leads to an eruption, i.e. we are dealing here
with an eruptive flare.

\begin{figure}
\includegraphics[scale=0.8]{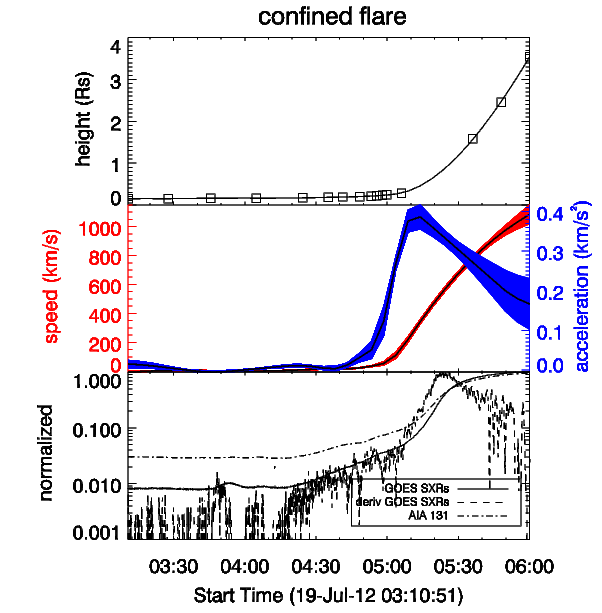}
\caption{Eruptive flare.  Top: temporal evolution of the flux
  rope height from measurements in the 335 \AA \, channel of AIA and
  LASCO C2 (squares; the last three data points correspond to LASCO
  measurements). The associated uncertainties are too small to be seen
  in this scale.  The heights resulting from the application of a
  smoothing cubic spline to the original measurements are shown with a
  solid black line. Middle: first (velocity) and second
  (acceleration) temporal derivatives of the smoothed height points
  (solid and dashed black lines respectively). Point-wise 1-$\sigma$
  uncertainties in the velocity (red) and acceleration (blue) from
  1000 Monte-Carlo simulations of the (assumed) Gaussian $h-t$
  uncertainties are also shown. Bottom: normalized light curves of the GOES 1-8 \AA \, flux (solid line),
  its temporal derivative (dashed line) and the 131 light curve of
  the box encapsulating the flux rope structure in Figure
  \ref{fig:conf}d (dashed-dotted line). 
} \label{fig:eruptht}
\end{figure}

As we did in the case of the confined flare, we deduce the kinematics of the
eruptive flux rope and compare them with the associated flare dynamics.  We
manually track the front of the erupting flux rope in the 335 images to
determine the $h-t$ profile (upper panel of Figure \ref{fig:eruptht}).  These measurements are complemented by the $h-t$ of the resulting
WL CME core observed in LASCO C2 (the last 3 datapoints in the upper
panel of Figure \ref{fig:eruptht}). An uncertainty of 5 pixels (0.043
\rs) is assigned to the LASCO measurements.  The same smoothing cubic
spline scheme used for the confined flare measurements, this time with
seven knots, is applied to the $h-t$ measurements to obtain a smoothed
$h-t$ profile. The first and second temporal derivatives provide the flux rope velocity and acceleration,
respectively. The derived speed and acceleration profiles along with
their point-wise 1-$\sigma$ uncertainties from 1000 Monte-Carlo
simulations of the (assumed) Gaussian $h-t$ uncertainties are
displayed in the middle panel of Figure \ref{fig:eruptht}.  Finally
the lower panel contains the GOES 1-8 \AA\ light
curve, its temporal derivative and the 131 light curve over the
box shown in Figure \ref{fig:conf}d.

Figure~\ref{fig:eruptht} leads to the following
remarks. The flux rope moves relatively slowly in the AIA FOV, reaching a
maximum speed of $\approx$ 100 \kms. The bulk of its acceleration
occurs beyond the AIA FOV where the flux rope speed exceeds 1000
\kms. The flux rope eruption leads to a fast CME.  The
flux rope acceleration rise consists of two phases: a gradual rise ($\approx$
04:40-04:55~UT) followed by an impulsive rise. The acceleration
reaches its peak at 5:10~UT.  Similarly, the associated flare (GOES
SXRs and 131 light-curves) exhibits a gradual rise ($\approx$
04:15-05:05~UT) followed by an impulsive rise phase ($\approx$05:05-05:25~UT). The flare
energy release rate (temporal derivative of GOES SXRs) evolves very similarly to the CME acceleration, exhibiting a gradual and impulsive phase and reaches its peak at around
05:25~UT.

The above findings suggest a tight correspondance between flare
heating and CME acceleration, as has been found already for the
majority of CMEs and also for the confined flare (see the detailed
discussion of CME-flare timings from statistical studies in Section
2.1).  However, (i) the impulsive rise of the acceleration starts
around 10 minutes \textit{before\/} the start of the impulsive rise of
the flare, and (ii) the peak of the acceleration occurs around 10
minutes \textit{before\/} the peak of the SXR temporal derivative. We
note here that the exact timings between the CME impulsive
acceleration and flare emissions are somehow uncertain because the
bulk of the CME impulsive acceleration takes place between the outer
edge of the AIA FOV and the inner edge of the C2 FOV, where
measurements are unavailable.  In the next sections, we focus on the
important implications of the time delays discussed above with respect
to the possible eruption trigger.

\section{3D context of the Event}
So far, we focused on the interpretation of the AIA observations
only. We believe that they provide a very convincing case for the
formation and subsequent eruption of a magnetic flux rope based on the
observed morphology, association with hot cusp-like loops, and
temperature evolution in the various AIA channels. Most of these
results are possible because of the fortuitous alignment of the flux rope
axis parallel to the AIA LOS thus providing an almost cartoon-like
view of the structure. At the same time, the single viewpoint AIA
observations cannot address the 3D configuration and low atmosphere
conectivity of the structure because these connections are hidden
behind the limb. 

Thankfully, we can take advantage of the EUVI-A 195
images which record a top view of the whole AR. Although SECCHI lacks
a dedicated hot channel like the AIA 131, the 195 passband includes
contributions from a Fe XXIV line at 192 \AA \, with peak temperature of 16 
MK. It can also
be compared rather directly with the AIA 193 images. Indeed, using the
AIA images as a guide, the flux rope can be barely identified as a faint loop
structure. The emission in the EUVI-A images, however, is dominated by
brightenings on either side of the NL, mostly lying along the southern
boundary, away from Earth (Figure~\ref{fig:sta}. We will focus on
these brigtnenings for our analysis because they play an important
role in deriving the 3D configuration of the flux rope as we will see
shortly. 

First, we recall the unsual AIA observations of the half-legs in
Figure~\ref{fig:conf}c-e. Both sides of the flare loops are clearly
visible while the flux rope core appears threaded by two distinct loop
bundles with a single footpoint originating at some distance from the
flaring loops. In other words, we have three footpoint clusters
somewhere south of the NL and a single footpoint cluster northward of
the NL. Why is that and what are those half-loops threading the flux rope?

\begin{figure}
\includegraphics[scale=0.9]{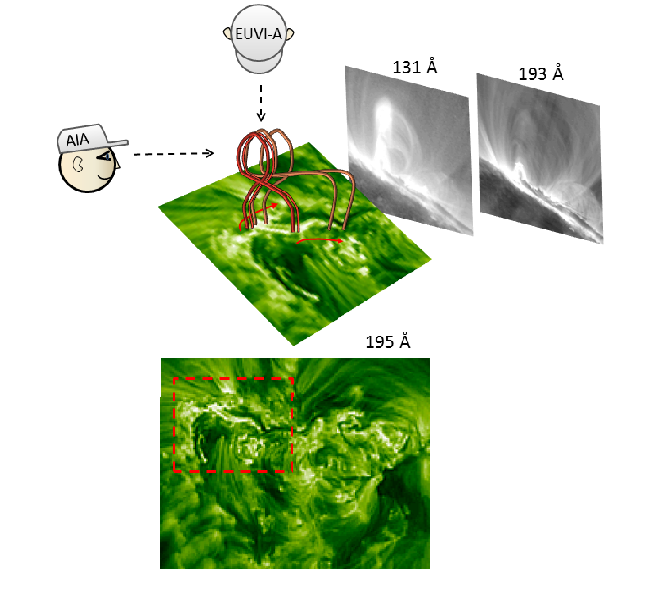}
\caption{The proposed 3D configuration of the flux rope. The flux rope structure
  is represented by a few hypothesized field lines (red and orange
  curves) that conform to the views from above (EUVI-A 195) and from
  the side (AIA 131 and 193). The red arrows show the progression
  of the brightenings along the surface and the red dotted-line box
  marks the area shown in detail at the center of the figure.} 
\label{fig:cartoon}
\end{figure}

The answer lies in the EUVI-A images taken at 22:10~UT
(Figure~\ref{fig:sta}c) almost simultaneously to the AIA 131 images in
Figure~\ref{fig:conf}c. The EUVI-A image shows three distinct
brightening areas. The most extended is the easternmost one which
corresponds to the flaring loop (and flux rope) seen from above. The other
two areas must be, therefore, the footpoints of the two loops bundles
that thread the flux rope. The separation probably explains the gap between
the two bundles as seen by AIA. The lack of any other significant
brightenings, north of the NL, means the loops originating in all
three southern locations must connect back along the easternmost
footpoint. The concentration of all these field lines and half-loop
appearance for the two core bundles can then be explained by a kinked
configuration where the kinked loops form the flux rope viewed in AIA with
their axis predominanlty parallel to the AIA LOS. We summarize the
resulting 3D configuration in Figure~\ref{fig:cartoon} where we plot
the AIA 131 and 193 images and the closest in time EUVI-A 195 image
as viewed from the two perspectives. We then draw our proposed
3D representation of a few field lines that is consistent with the
observations from the two viewpoints. \cite{pvk} deduced a similar
configuration in another event based on detailed 3D analysis although they lacked the high cadence
and temperature coverage of the AIA instrument. 
 \cite[][]{ji03,alex06,kink}
reported on very clearly kinked prominences which were also  failed
eruptions. Our interpretation in Figure~\ref{fig:cartoon}, therefore,
does not seem unreasonable. It does imply, however, that a kinked flux rope
can survive for quite some time (at least 6 hours) before erupting.

\section{Discussion}
In this work, we present the first {\it direct} unambiguous evidence
of a pre-existing flux rope involved in a fast CME eruption. Thanks to
 EUV observations in many passbands and from two viewpoints,
we can follow the temporal and spatial evolution of the system in great detail. To recap,
we first present a brief event timeline with the approximative
times of the most important aspects of our observations, with
the entire sequnece running from 22:00~UT on 18 July until
05:36 on 19 July 2012. 
\begin{itemize}
\item {\bf 22:00-22:30.} A magnetic flux rope is formed during a confined flare.
\item {\bf 22:30-02:10.}  The flux rope plasma cools appearing sequentially
in EUV channels with peak temperatures ranging from flaring to
transition region conditions.
\item {\bf 22:30-04:00.} The flux rope
and overlying coronal structures undergo a phase
of slow quasi-static rise (speed 0.5-2 \kms) and expansion.
\item  {\bf  02:47-03:57.} a hot cusp
loop structure and new legs threading the flux rope
appear in the 131 channel only.
\item {\bf 04:45-05:36.} the flux rope
begins to impulsively accelerate and a WL CME appears.  
\end{itemize}

\subsection{Event Sequence Scenario}
In the previous sections, we split the flux rope evolution in three phases
(formation, quasi-static rise and expansion, and eruption). We now
incorporate them into a coherent physical scenario which involves
ideal and non-ideal physical processes.

\begin{enumerate}
\item {\it Flux rope formation} The observations supply strong
  evidence that the flux rope is formed via magnetic reconnection during the
  confined flare. The evidence
  includes the simultaneous formation of cusp-like loops below the flux rope
  with flare temperatures, the tight synchronization between the flare
  energy release and the evolution of the flux rope size, and the
  distribution of flare brightenings along the NL. The high initial
  altitude and size of the flux rope points to a {\it coronal} flux rope.  Overall,
  the observations point to the conversion of arcade magnetic fields
  via successive reconnections to flux rope helical fields. This mechanism of flux rope
  formation is addressed in a number of theoretical, modeling, and
  observational works
  \citep[e.g.,][]{balle89,moore92,low96,anti1999,amari2000,lin04,lynch08,vra08,green09,geor11}
  which our observations are now verifying.

\item {\it Slow quasi-static rise} It is well established that slow
  photospheric footpoint motions leading to shearing and/or twisting
  can drive the slow rise and inflation of coronal structures. This
  mechanism can be purely ideal, i.e. not associated with magnetic
  reconnection, and can explain, in principle, the observed
  quasi-static rise of the flux rope and overlying corona.  However,
  we observe the formation of a cusp structure along with "half-loop"
  legs threading the flux rope core starting at around 02:47~UT during this phase.
  This clearly suggests
  that low reconnection-rate phenomena are taking place.
  Moreover, the confined flare and coronal flux rope exhibit an "X"-type
  topology.  We therefore suggest that small magnitude tether-cutting
  reconnections may be transforming arcade fields into flux rope fields
  causing the growth and slow rise of the initial flux rope. This process
  could be essentialy the same process responsible for the flux rope
  formation in the previous paragraph but occurring with lower
  magnitudes of the related phenomena. The cause of the slow
  rise observed before 02:47~UT is less certain.

\item {\it Magnetic seed and trigger of the fast CME} The 'seed' for
  the fast CME on July 19 was the destabilization of the pre-existing
  flux rope formed almost seven hours prior to the onset of the impulsive
  acceleration phase of the CME.  The delay in the onset of the SXR
  impulsive phase and in the peak of the SXR temporal derivative with
  respect to the onset and peak of the flux rope acceleration
  (see Figure \ref{fig:eruptht} and corresponding discussion) points to a
  flux-rope (i.e. {\it ideal}) instability for the trigger of the
  eruption. In other words the strong acceleration of the plasma
  starts {\it before} the strong flare heating. The kink
  \citep[e.g.,][]{kink} and torus instabilities \citep[e.g.,][]{torus}
  are two commonly invoked flux rope instabilities for the trigering
  of CMEs.  However, the lack of hard
  evidence of flux rope rotation during the eruption, which could have
  served as an indication of
  kink instability, suggest we can probably exclude this instability as the 
  trigger of the eruption.  On the other hand, the slow and long
  duration quasi-static rise of the flux rope could have lifted it to
  altitudes where the overlying constraining magnetic
  field gradients are stronger, thus facilitating the onset of the torus instability as
  discussed in a number of series of recent MHD modeling
  investigations \citep[e.g.,][]{aula10,fan10,sav12}.
  Mechanical loss of equilibrium of a flux rope once it
  reaches a critical height 
\citep[e.g.,][]{forb91,forbes00,vra08}
is another possible trigger for the
  observed eruption.

\end{enumerate}

To summarize, the above scenario contains both ideal and non-ideal
processes with the ideal processes possibly triggering the eruption
while the non-ideal processes are responsible for the flux rope formation and
its subsequent slow quasi-static rise.

\subsection{Implications for CME Initiation}

For the first time, we have confirmation that truly pre-existing hot coronal
flux ropes exist and can be long-lived. Note that previous
observations detected a hot flux rope forming only a few minutes
before the onset of the associated eruption
\citep{cheng11,zhang12}. The very significant span of almost seven
hours between the flux rope formation and its eruption underlines the
importance of studying the long-term evolution of ARs and to not only
concentrate on the immediate time around the eruption. We believe that
we would have reached significantly different, possibly
erroneous, conclusions regarding the onset of the CME on July 19 had
we focused only on the events surrounding the CME onset.  We are also in
the position to explain the rarity of past detections of pre-existing
flux ropes: (1) Lack of high cadence EUV imaging observations in flare
temperatures prior to AIA.  (2) Small spatial (horizontal and
vertical) and temporal scales of the flux rope requiring high spatial
resolution. (3) Favorable line of sight orientation with respect to
the flux rope axis. The initially small spatial scales require that
the flux rope is situated right at the limb so that it would be
vissible before its destabilization and rapid ascent. All of these
requirements were met by our observations.

Our study also supplies tight constraints for models of flux rope formation
and eruption.  The flux rope was formed in a period of $\approx$ 20 minutes.
The flux rope was a rather small structure during its formation. Its initial
(final)  minor and major axis had  lengths of $\approx$ 4.7
(22.2) and  9 (41.5) Mm, respectively.  Moreover, the flux rope evolved
within a small range of heights, $\approx$ 80 Mm at the end of the
confined flare and 138 Mm at the start of the impulsive CME
acceleration.  The period of slow quasi-static rise lasted for almost
seven hours.  Therefore, succssful flux rope models need to reproduce rapidly
formed small and low-lying coronal flux ropes with long quasi-static
intervals before eruption.

The present work suggests that the events taking
place during confined flares may play an important role
in, at least some, solar eruptions. The magnetic field
reconfiguration associated with the confined flare in this event
could be considered as a catalyst for the sequence of events leading
to the CME.  It did not only create the flux rope which was the seed for the
CME but it also re-configured the field by the formation and initial
rise of the rope by setting up a topology (i.e., X-point) which favors
(coronal) magnetic reconnection.  We note here that large numbers of
small magnitude confined flares may take place in the source AR
before, and after, major solar eruptions and eruptive flares (e.g., M
and X class). Such small magnitude events may release only part of the
accumulated free energy in the form of radiation while some portion of
the free energy ends up as magnetic energy of a flux rope. 
A similar picture of flux rope formation by small/confined flares
was theoretically formulated by \citet{low96}.
Indeed,
inspection of the evolution of our AR over the long interval of 2012
July 17-20 gives some hints of a {\it homologous} behavior with a few
confined flares giving rise to flux-rope like structures. Evidently,
the LOS orientation was not as optimal as for the event on July 18.
In addition, all flaring and eruptive activity seems to originate from
around the same location along the NL.  Obviously our hypothesis about
the overall role of confined flares in the formation and initiation of
CMEs requires further testing against more observations.

Regarding future instrumentation for addressing the CME initiation
problem, our study makes it clear that a coronagraphic capability in
flare EUV lines withing the inner corona ($<1.5$ \rs) may be an important
element of a mission on Solar Eruptive Events. Such observations will
allow us to trace the hot flux rope as far as possible in the low corona, and
avoid the strong footpoint emissions and image saturation and
diffraction effects from the associated flares which may mask the
faint flux rope emissions in current telescopes. An obvious logical extension
of this study is a survey of the AIA database for more eruptive events
to access how common hot pre-existing (or otherwise) flux ropes are
and how and when do they form and eventually erupt.

\begin{acknowledgements}
The AIA data used here are courtesy of \textit{SDO\/} (NASA) and the
AIA consortium.  We thank the AIA team for the easy access to
calibrated data.  We thank the referee for useful comments
on the manuscript
and S. K. Antiochos, J. T. Karpen, S. Lukin and J. Zhang for
useful discussions.
The SECCHI data are courtesy of \textit{STEREO\/}
and the SECCHI consortium. 
This research has been partly co-financed 
by the European Union (European Social Fund – ESF) and Greek 
national funds through the Operational Program "Education and Lifelong Learning" of the National 
Strategic Reference Framework (NSRF) - Research Funding Program:
Thales. Investing in knowledge society through the European Social
Fund. S.P. acknowledges support from an FP7 Marie Curie 
Grant (FP7-PEOPLE-2010-RG/268288).
A.V. is supported by NASA contract S-136361-Y to 
the Naval Research Laboratory.  The work of G.S. was partly funded by NASA contract NNX11AD40G.
\end{acknowledgements}

\clearpage

\end{document}